\documentclass[
    ,final            
  ]
  {aipproc}

\layoutstyle{8x11double}

\begin{document}

\title{Why are massive O-rich AGB stars in our Galaxy not S-stars?}


\author{D.A. Garc\'{\i}a-Hern\'andez}{address={European Space Astronomy Centre
(ESAC), ESA. Apdo. 50727. 28080 Madrid. Spain}
}

\author{P. Garc\'{\i}a-Lario}{address={European Space Astronomy Centre (ESAC), ESA. Apdo. 50727. 28080 Madrid. Spain}
}

\author{B. Plez}{address={GRAAL, CNRS UMR 5024, Universit\'e de
Montpellier 2, 34095 Montpellier Cedex 5, France}
}

\author{A. Manchado}{address={Instituto de Astrof\'{\i}sica
de Canarias, La Laguna, E$-$38200, Tenerife, Spain}
}

\author{F. D'Antona}{address={Osservatorio Astronomico di Roma, via Frascati 33, 00040 MontePorzio Catone,
Italy }
}

\author{J. Lub}{address={Sterrewacht Leiden, Niels Bohrweg
2, 2333 RA Leiden, The Netherlands}
}

\author{H. Habing}{address={Sterrewacht Leiden, Niels Bohrweg
2, 2333 RA Leiden, The Netherlands}
}


\begin{abstract}
We present the main results derived from a chemical analysis carried out on a
large sample of galactic O-rich AGB stars using high resolution optical
spectroscopy (R$\sim$40,000--50,000) with the intention of studying their
lithium abundances and/or possible s-process element enrichment. Our chemical 
analysis shows that some stars are lithium overabundant while others are not.
The observed lithium overabundances are interpreted as a clear signature of the
activation of the so-called ``Hot Bottom Burning'' (HBB) process in massive
galactic O-rich AGB stars, as predicted by the models.  However, these stars do
not show the zirconium enhancement (taken as a representative for the s-process
element enrichment) associated to the 3$^{rd}$ dredge-up phase following
thermal pulses. Our results  suggest that the more massive O-rich AGB stars in
our Galaxy  behave differently from those in the Magellanic Clouds, which are
both Li- and s-process-rich (S-type stars). Reasons for this unexpected result
are discussed. We conclude that metallicity is probably the main responsible
for the differences observed and suggest that it may play a more important role
than generally assumed in the chemical evolution of AGB stars.

\end{abstract}

\maketitle


\section{Introduction}
The Asymptotic Giant Branch (AGB) is formed by stars with initial masses in the
range between 0.8 and 8 M$_{\odot}$ in a late stage of their evolution. During
most of the time H burning is the main source of energy for the AGB star but,
occasionally, the inner He shell ignites in a ``thermal pulse''  and,
eventually, the byproducts of He burning may reach the outer layers of the
atmosphere (the so-called 3$^{rd}$ dredge-up). Thus, AGB stars, originally
O-rich, can turn into C-rich AGB stars (C/O $>$ 1) after a few thermal pulses.
Another important characteristic of AGB stars is the presence of neutron-rich
elements (s-elements like Rb, Zr, Ba, Tc, Nd, etc.) in their atmospheres which
are the consequence of the slow-neutron captures produced during the thermal
pulsing phase.  According to the most recent models two major neutron sources
can operate in AGB stars depending on the stellar mass. The
$^{13}$C($\alpha$,n)$^{16}$O reaction is the preferred neutron source for
masses around 1$-$3 M$_{\odot}$ while for intermediate mass stars (M $>$ 3
M$_\odot$) the neutrons are thought to be mainly released by
$^{22}$Ne($\alpha$,n)$^{25}$Mg (see e.g. Lattanzio \& Lugaro 2005 for a recent
review).

In the case of the more massive O-rich AGB stars (M $>$ 4 M$_\odot$), the
convective envelope can penetrate the H-burning shell activating the so-called
``Hot bottom burning'' (HBB) process. HBB takes place when the temperature at
the base of the convective envelope is hot enough (T $\geq$ 2$\times$10$^{7}$
K) and $^{12}$C can be converted into $^{13}$C and $^{14}$N through the CN
cycle (Sackmann \& Boothroyd 1992). HBB models (e.g. Mazzitelli, D'Antona \&
Ventura 1999, hereafter MDV99) predict also the production of $^{7}$Li by the
chain $^{3}$He($\alpha$,$\gamma$)$^{7}$Be (e$^{-}$,$v$)$^{7}$Li, through the
so-called ``$^{7}$Be transport mechanism'' (Cameron \& Fowler 1971). One of the
predictions of these models is that Li should be detectable, at least for some
time, on the stellar surface. 

The HBB activation in massive O-rich AGB stars is supported by studies of AGB
stars in the Magellanic Clouds (hereafter, MCs) (e.g. Plez, Smith \& Lambert
1993). The detection of strong Li overabundances together with strong s-element
enhancement in these massive (and luminous) AGB stars is the signature that
they are indeed HBB stars which have undergone a series of thermal pulses and
dredge-up episodes in their recent past. 

In our own Galaxy, only a handful of Li-rich stars have been found so far (e.g.
Abia et al. 1993). Most of them are low mass C-rich AGB stars (e.g. Abia \&
Isern 2000) and intermediate mass S- and SC-stars (e.g. Abia \& Wallerstein
1998) and not O-rich M-type stars. However, HBB is expected to be active in the
most massive (and luminous) AGB stars (from $\sim$4 to 7 M$_{\odot}$ according
to MDV99 HBB models), which might not be C-rich, but O-rich. The best
candidates are the so-called {\it OH/IR stars},  luminous O-rich AGB stars
extremely bright in the infrared, showing a characteristic double-peaked OH
maser emission at 1612 MHz. These stars are also known to be very long period
variables (LPVs), sometimes  with periods of more than 500 days and large
amplitudes of up to 2 bolometric magnitudes. However, they experience very
strong mass loss rates (up to several times 10$^{-5}$ M$_{\odot}$yr$^{-1}$) and
most of them are usually heavily obscured at this stage by thick circumstellar
envelopes, making optical observations very difficult. Thus, no information 
exists yet on their Li abundances and/or possible s-process element
enrichment. 


\section{Observations and results}
A large sample (102) of long-period (300$-$1000 days), large amplitude
variability (up to 8$-$10 magnitudes in the V band), late-type ($>$ M5) O-rich
AGB stars displaying OH maser emission with a wide range of expansion
velocities (from just a few km s$^{-1}$ to more than 20 km s$^{-1}$) was
carefully selected. Stars were included in the sample if satisfying at least
one of the above criteria and ideally as many of them as possible, which
guarantees that they are actually massive stars. Consistently, stars in the
sample were mainly members of the galactic disk population and displayed strong
IR excesses detected by IRAS.

High-resolution optical echelle spectra (R$\sim$40,000--50,000) were obtained
for all stars in the sample during several observing periods in 1996--1997. The
full log of the spectroscopic observations is shown in Table 1, including more
detailed information on the observations. The two-dimensional frames containing
the echelle spectra were reduced to single-order one-dimensional spectra using
the standard {\sc echelle} software package as implemented in
IRAF\footnote{Image Reduction and Analysis Facility (IRAF) software is
distributed by the National Optical Astronomy Observatories, which is operated
by the Association of Universities for Research in Astronomy, Inc., under
cooperative agreement with the National Science Foundation.}. Because of the
very red colours of the sources observed, the S/N ratios achieved in the
reduced spectra can strongly vary from the blue to the red orders (10-20 at
$\sim$6000 \AA~while $>$100 at $\sim$8000 \AA).

\begin{table}
\footnotesize
\begin{tabular}{cccc}
\hline
    \tablehead{1}{c}{b}{Set-up\\}
  & \tablehead{1}{c}{b}{Date\\}
  & \tablehead{1}{c}{b}{$\Delta\lambda$\\(\AA/pix)}
  & \tablehead{1}{c}{b}{Range\\(\AA)}  \\
\hline
4.2m WHT/UES    &August 1996   &0.065  & 5300-9400\\
4.2m WHT/UES    &June 1997     &0.065  & 4700-10300\\
4.2m WHT/UES    &August 1997   &0.065  & 4700-10300\\
3.6m ESO/CASPEC &February 1997 &0.085  & 6000-8200\\
\hline
\end{tabular}
\caption{Log of the spectroscopic observations}
\label{tab:a}
\end{table}

We detected the presence of the Li I resonance line at 6708 \AA~in 25\% of the
sources in the sample with a wide variety of strengths, while we did not find
any signature of this line in 31\% of the stars. The remaining 44\% were
heavily obscured by their thick circumstellar envelopes and they were too
red/not found at optical wavelengths. In general, all stars (with or without
lithium) show extremely red spectra with the flux level falling dramatically at
wavelengths shorter than 6000 \AA. In addition, the spectra are severely
dominated by strong molecular bands mainly due to titanium oxide (TiO), as a
consequence of the very low temperature and the O-rich nature of these stars.
Interestingly, the bandheads of ZrO seem to be absent in all spectra. This is
shown in Figure 1 where we show the spectral region around the ZrO bandheads at
6474 and 6495 \AA. These ZrO bandheads (as well as those corresponding to other
s-element oxides such as LaO or YO) are very strong in galactic S-stars and in
massive MC AGB stars.


\begin{figure}
\centering
\includegraphics[width=9cm,height=15cm,angle=-90]{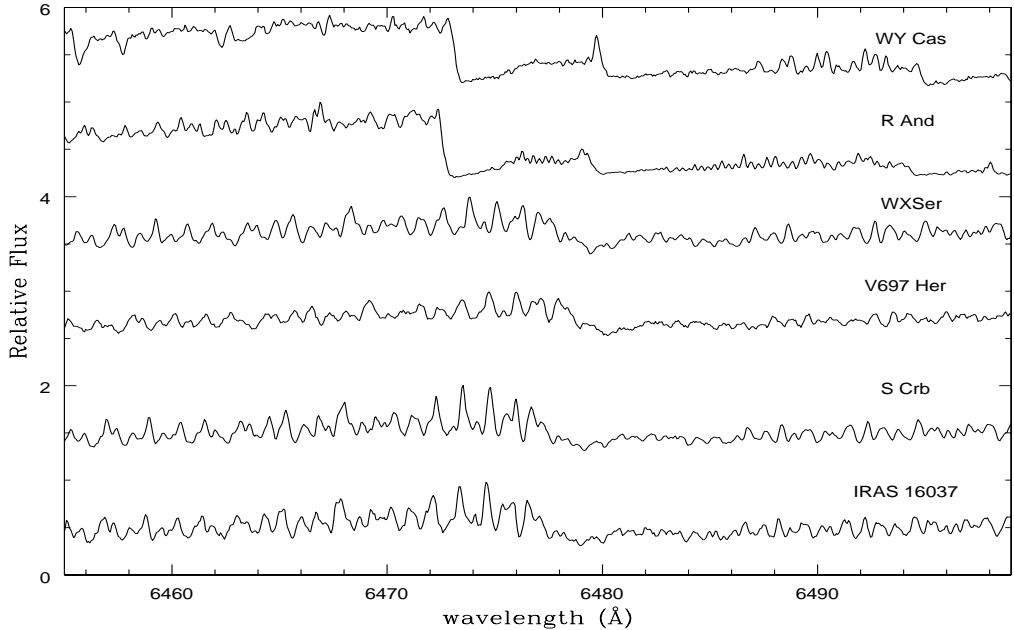}
\caption{High resolution optical spectra of sample stars displaying 
the lack of the ZrO absorption bands at 6474 and 6495 \AA~compared with two
galactic S-stars (WY Cas and R And). WX Ser (IRAS 15255$+$1944) and V697 Her
(IRAS 16260$+$3454) are Li-detected while S CrB (IRAS 15193$+$3132) and IRAS
16037$+$4218 are Li non-detected. The absorption band
at $\sim$6480 \AA~corresponds to the TiO molecule.}
\end{figure}


\section{Chemical analysis}
Our analysis combines state-of-the-art line blanketed model atmospheres and
synthetic spectroscopy with extensive linelists. We have used the spherically
symmetric, LTE, hydrostatic `MARCS' model atmospheres for cool stars and the
`TURBOSPECTRUM' spectral synthesis code (Alvarez \& Plez 1998) to derive the Li
and Zr (taken as representative of all other s-process elements) abundances in
those stars for which an optical spectrum could be obtained. 

From an exhaustive study of the influence of the variations of the fundamental
stellar parameters \textit{(e.g. T$_{eff}$, log g, M, z, $\xi$, C/O, etc.)} on
the synthetic spectra and from our knowledge of the main characteristics of our
stars we obtained the most adequate initial set of parameters as well as their
plausible range of variation, and we constructed a grid of MARCS model spectra.

Thus, we first determined by $\chi$$^{2}$ minimisation which of the spectra
from our grid of models provided the best fit to the observations in the
6670$-$6730 \AA~and the 6455$-$6499 \AA~spectral regions. The goal was to fit
the overall shape of the spectra including the TiO bandheads, which are very
sensitive to variations in the effective temperature. Then, the Li and Zr
abundances were derived by fitting the Li I resonance line at $\sim$6708
\AA~and the ZrO molecular bands at 6474 \AA~and 6495 \AA, respectively. As an
example, the best fit in the 6670$-$6730 \AA~spectral region around the Li I
line is presented in Figure 2 for the star IRAS 11081$-$4203.

\begin{figure}
\centering
\includegraphics[width=9cm,height=15cm,angle=-90]{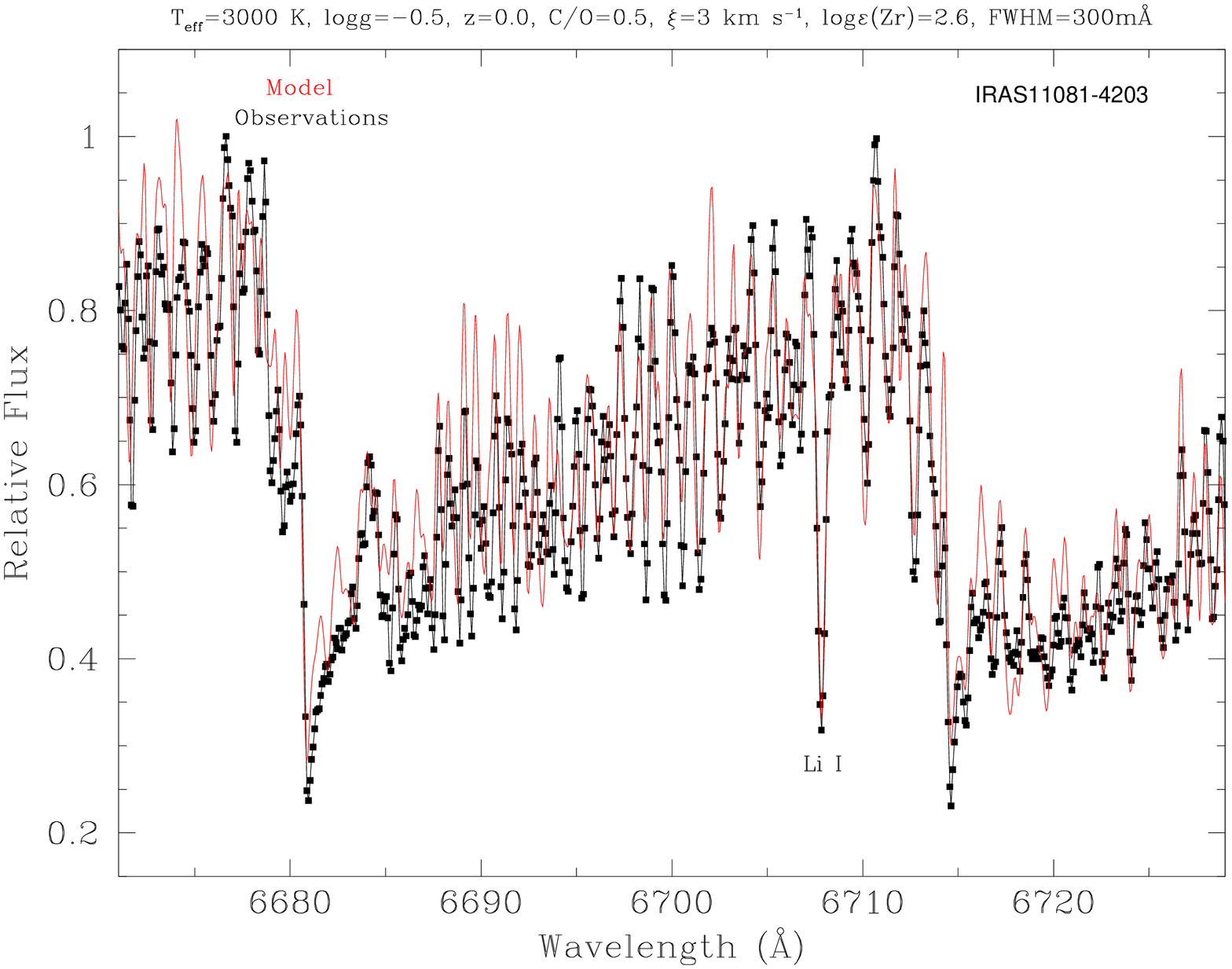}
\caption{Best model fit and observed spectrum in the region 6670$-$6730 \AA~for
the star IRAS 11081$-$4203. The \textit{T$_{eff}$} and Li abundance derived
from this spectrum was 3000 K and \textit{log $\varepsilon$(Li)}=1.3,
respectively. The parameters of the best model atmosphere fit are
indicated in the top label.}
\end{figure}


\section{Li and Zr abundances}
Our chemical abundance analysis shows that half of the stars show Li
overabundances in the range \textit{log $\varepsilon$(Li)}$\sim$0.5 and
3.0\footnote{Li abundance in the scale 12$+$\textit{log N(Li)}. Note that the
uncertainty in the Li abundances derived is estimated to be of the order of
0.4$-$0.6 dex. This error reflects mostly the sensitivity of the derived
abundances to changes in the atmospheric parameters taken for the modelling.}.
A very similar range of Li overabundances is found in the massive O-rich AGB
stars studied in the MCs (e.g. Plez, Smith \& Lambert 1993). The Li
overabundances observed are interpreted as a signature of the activation of the
so-called ``Hot Bottom Burning'' (HBB), confirming that they are massive AGB
stars (M $>$ 4 M$_\odot$ according to MDV99 HBB models).

However, the non-detection of the ZrO molecular bands at 6474 \AA~and 6495
\AA~in any of the stars analysed imposed severe upper limits to the zirconium
abundance ([Zr/Fe]$<$0.0$-$0.25 for \textit{T$_{eff}$} $\geq$ 3000 K and
[Zr/Fe]$<$0.25$-$0.50 for \textit{T$_{eff}$} $<$ 3000 K). If the Zr
enhancement is taken as a representative for the s-process enrichment, our
results indicate that the massive AGB stars in our Galaxy are not S-stars.

\section{Comparison with the Magellanic Clouds}
In contrast with their galactic analogues, the more massive AGB stars in the
MCs are O-rich stars showing s-process elements enhancement (S-stars). In
addition, a higher proportion of them ($\sim$80\% compared to $\sim$50\% in our
Galaxy) shows also Li enhancement. The Li enhancement indicates that they are
also HBB stars but, why are these stars also enriched in s-process elements? 

The answer to this question must be related to the different metallicity of
the stars in the MCs with respect to our Galaxy. Actually, theoretical models
predict a higher efficiency of the dredge-up in low metallicity atmospheres
(e.g. Herwig 2004) with respect to those with solar metallicity (e.g. Lugaro
et al. 2003). 

In addition, there is an increasing observational evidence that lower
metallicity environments are also less favourable to dust production, as it is
suggested by the very small number of heavily obscured AGB stars in the MCs
(e.g. Groenewegen et al. 2000). This is supported by the lower
dust-to-gas ratios derived by van Loon (2000) in the few obscured MC AGB stars 
for which this analysis has been made. If mass loss is driven by radiation
pressure on the dust grains, this might be less efficient with decreasing
metallicity (Willson 2000). In that case, longer AGB lifetimes would be
expected, which could increase the chance of nuclear-processed material to
reach the stellar surface. The slow evolution predicted for AGB stars in the
MCs as a consequence of the less efficient mass loss leaves time for more
thermal pulses to occur during the AGB lifetime and, therefore, a more
effective dredge-up of s-process elements to the surface can be expected before
the envelope is completely gone at the end of the AGB. This would explain why
even the more massive stars in the MCs show a strong s-process enrichment in
contrast to their galactic counterparts. In our Galaxy the only AGB stars
showing a similar overabundance in s-process elements seem to be the result of
the evolution of low- to intermediate-mass stars (M $<$ 1.5$-$2.0 M$_\odot$),
while no or very litle s-process enhancement is observed in galactic AGB stars
with higher main sequence masses. 

Finally, the lower critical mass needed to develop HBB (e.g. M $>$ 3 M$_\odot$
at the metallicity of the LMC, compared to the $\sim$4 M$_\odot$ limit in our
Galaxy) would favour the simultaneous detection of s-process elements and Li
enrichment in a larger number of AGB stars in the MCs, as it is actually
observed. In contrast to their MC counterparts, Li-rich massive AGB stars in
our Galaxy would evolve so rapidly (because of the strong mass loss) that there
is no time for a significant enhancement in s-process elements.

In summary, our results suggest that the dramatically different abundance
pattern found in AGB stars belonging to the MCs and to our Galaxy can be
explained in terms of the different metallicity conditions under which these
stars evolved. This is the first observatinal evidence that the chemical
evolution during the AGB could be strongly modulated by metallicity. A
complete description and discussion of these results as well as their
evolutionary consequences will be given in Garc\'{\i}a-Hern\'andez et al.
(2005, in preparation).





\bibliographystyle{aipproc}   





\end{document}